\documentclass[12pt,dvips]{article}
\usepackage{rotating}
\usepackage{epsfig}
\usepackage{graphicx}
\usepackage{color}
\usepackage{cite}

\def\be{\begin {equation}}
\def\ee{\end {equation}}
\def\beqa{\begin {eqnarray}}
\def\eqa{\end {eqnarray}}
\def\ba*{\begin {eqnarray*}}
\def\ea*{\end {eqnarray*}}
\def\m{\mathrm}

\begin{document}

\def\thefootnote{\fnsymbol{footnote}}

\begin{flushright}
{\tt INFN/TC\_11/9}
{~}\\
\vspace{0.5cm}
January 2012
\end{flushright}

\begin{center}
{\bf {\Large Electrostatic Storage Ring}}
\end{center}

\medskip

\begin{center}{\large
{\bf Mario~Conte}
}
\end{center}

\begin{center}
{\em Dipartimento di Fisica dell'Universit\`a di Genova and
  INFN-Sezione di Genova, 
  Via Dodecaneso 33, 16146 Genova, Italy.}\\[0.2cm]


\end{center}

\bigskip

\centerline{\bf ABSTRACT}

\noindent  
In the trial \cite{BNL} of measuring the proton electric moment, storage rings with electrostatic lattice have been considered. 
Here an overview is given about the main parameters regarding such a kind of focusing. Beyond confirming all the issues 
regarding this subject, a non-null element $M_{31}$ is introduced in all the $3\times 3$ matrices which deal with the vector 
$(x,x',\Delta p/p)$ and its role is discussed.

\medskip
\noindent

\newpage

\section {Introduction}
Let us define \cite{Reiser1,Lawson,Ferrando} the basic equation expressing the radial balance of forces
\be
  \frac{d}{dt}(\gamma m \dot{x}) = \frac{\gamma m\beta^2 c^2}{R} + qE_x  = f                     \label {radeq}
\ee
where
\beqa
       R & = & \rho + x                \nonumber   \\
  \gamma & = & \gamma_0 + \Delta\gamma \nonumber   \\
  \beta  & = & \beta_0 + \Delta\beta   \nonumber   \\
     E_x & \simeq & [E_x]_{x=0}+\left(\frac{\partial E_x}{\partial x}\right)_{x=0} =                      
                                                 E_{x_0}\left(1-n\frac{x}{\rho}\right)     \nonumber  \\
       n & = & -\frac{\rho}{E_{x_0}}\left(\frac{\partial E_x}{\partial x}\right)_0         \label {flindex}
\eqa
with the label ``0'' referring to that trajectory where $f$ of Eq. (\ref{radeq}) is null and 

\be 
      qE_{x_0} = - \frac {\gamma_0\beta_0^2 mc^2}{\rho}                                           \label {eqeq} 
\ee
Hence we can rewrite the right side of Eq. (\ref{radeq}) as follows:
{\small
\beqa
  f & = & \frac {\gamma_0\beta_0^2 mc^2}{\rho}\;
          \frac {\left(1+\frac{\Delta\gamma}{\gamma_0}\right)\left(1+\frac{\Delta\beta}{\beta_0}\right)^2}
          {1+ \displaystyle \frac{x}{\rho}} + qE_{x_0}\left(1-n\frac{x}{\rho}\right)                        \nonumber   \\
  f & \simeq & -qE_{x_0}\left(1+\frac{\Delta\gamma}{\gamma_0}\right)\left(1+2\frac{\Delta\beta}{\beta_0}\right)
                   \left({1-\frac{x}{\rho}}\right)+qE_{x_0}-qE_{x_0}n\frac{x}{\rho}           \nonumber   \\
  f & \simeq & -qE_{x_0}\left(1+\frac{\Delta\gamma}{\gamma_0}+2\frac{\Delta\beta}{\beta_0}-\frac{x}{\rho}\right)+                     
  qE_{x_0}-qE_{x_0}n\frac{x}{\rho}                                               \nonumber   \\      
  f & \simeq & -qE_{x_0}\left[\frac{\Delta\gamma}{\gamma_0}\left(1+\frac{2}{\gamma_0^2-1}\right)-
                                          (1-n)\frac{x}{\rho}\right]                        \label {f1}
\eqa }
since
$$   \frac{\Delta\beta}{\beta_0} = \frac{1}{\gamma_0^2-1}\frac{\Delta\gamma}{\gamma_0}.  
$$
Elaborating further the relativistic expressions, we have:
$$
   1+\frac{2}{\gamma_0^2-1} = \frac{\gamma_0^2+1}{\gamma_0^2-1} = \frac {1+1/\gamma_0^2}{1-1/\gamma_0^2} =
   \frac {2-\beta_0^2}{\beta_0^2} ~~~~~~ {\m since} ~ 1/\gamma_0^2 = 1-\beta_0^2
$$
which transforms Eq. (\ref{f1}) into
\be
   f = -qE_{x_0}\left[\frac {2-\beta_0^2}{\beta_0^2}\frac{\Delta\gamma}{\gamma_0} - 
                                             (1-n)\frac{x}{\rho}\right]                        \label {f2}
\ee            
Then, considering that a small radial displacement $x$ of the charged particle, generated by the electric field, 
implies a small variation of the total energy
\be
     qE_{x_0}x \Delta (U_{\m tot}) = mc^2\Delta\gamma = qE_{x_0}x  ~~~~~~~~ {\m or} ~~~~~~~~ 
    \frac{\Delta\gamma}{\gamma_0} = \frac{qE_{x_0}}{\gamma_0 mc^2}x = -\beta_0^2 \frac {x}{\rho}   \label {engain}
\ee
having taken into account Eq. (\ref{eqeq}). Therefore, Eq. (\ref{f2}) becomes

\be 
    f = qE_{x_0}(3-\beta_0^2-n)\frac{x}{\rho}                                          \label {f3}
\ee

\section {Motion equations}
In order to settle the motion equations, we need to know the energy of the circulating particles. In fact, developing the left side 
of Eq. {\ref{radeq}}, we obtain
\be
    \frac{d}{dt}(\gamma m \dot{x}) = \gamma m \frac {d^2 x}{dt^2} + m\dot{x}\frac{d\gamma}{dt} \simeq  
                                  \gamma m \frac {d^2 x}{dt^2}                              \label {apprstudy} 
\ee
where the approximation is correct if we are in the non relativistic \cite{Ferrando} regime. Otherwise, in order to discuss this 
topic we need to introduce some realistic value of the beam energy such as, for instance, the protons ``magic'' energy; i.e. a 
Lorentz factor
\be
         \gamma_{\m m} = \sqrt{1+\frac{1}{a}} = 1.248                              \label {magigam}
\ee
where $a=1.793$ is the proton anomaly. Consequently we have:
\ba* 
   \beta_{\m m}  & = & 0.598                                 \\
   p_{\m m} & = & \frac {mc}{\sqrt{a}} =  0.701 ~ \m{GeV/c}        \\
   U_{\m m} & = & \gamma_{\m m}\,mc^2 =  1.171 ~ \m{GeV}    \\
   W_{\m m} & = & U_{\m m}-mc^2 = 0.233 ~ \m{GeV}           \\
   m & = & \m{proton ~ mass} = 0.938 ~ \m{GeV/c^2}
\ea*
and
\be
  B_{\m{eq}} = \frac {E_{\m{rad}}}{\beta c}, ~~~ \m{for} ~~~ E_{\m{rad}} = 1.5 \times 10^7 ~ 
                                                      {\m V}{\m m}^{-1} = 150 ~ \m{kV/cm}     \label {E-bend}
\ee
or
\be 
  B_{\m{eq}} = 8.37 \times 10^{-2} ~ \m{Tesla}, ~ \m{which~yields~a~bending~radius} ~
                                            \rho = \frac{p}{eB_{\m{eq}}} = 28 ~ {\m m}          \label {r-bend}
\ee
At this stage we can elaborate Eq. (\ref{apprstudy}) as follows:

\small
\be
   \frac{d}{dt}(\gamma m \dot{x}) = \gamma m\ddot{x} + m \dot{x} \frac{d\gamma}{dt} = 
       \gamma m \left(\ddot{x} + \frac{qE_{\m x0}}{U_{\m tot}} \dot{x}^2\right)                   \label {compar1}     
\ee 
since, due to Eq. (\ref{engain}), we have:
{\small
$$
   \frac{d\gamma}{dt} = \frac{1}{mc^2}\frac{dU_{\m tot}}{dt} = 
   \frac{1}{mc^2}\frac{d}{dt}(qE_{\m x0}x) = \frac{qE_{\m x0}}{mc^2} \dot{x}
$$ }  
We can reasonably suppose that the solutions should be of the sinusoidal type, i.e. $x(t)\simeq \sin(\omega_{\m H}t)$ and 
$y(t)\simeq \sin(\omega_{\m V}t)$, and therefore we have
{\small
$$
    \ddot{x} = -(\omega_{\m H})^2 X \sin(\omega_{\m H}t)  ~~~~ {\m and} ~~~~
   (\dot{x})^2 = (\omega_{\m H}X)^2 \cos^2(\omega_{\m H}t)
$$ }
Comparing the maximum values of the two terms within brackets of Eq. (\ref{compar1}), we find that the acceleration
$\ddot{x}$ is much bigger than the term containing $\dot{x}^2$ if
$$   
   X \ll \frac {U_{\m tot}}{qE_{\m x0}} = 78.24 ~ {\m m}
$$
a condition absolutely fulfilled since $X$ will be of the order of a few millimeters. Therefore we may confirm the approximate 
version of Eq. (\ref{apprstudy})
\be
  \frac{d}{dt}(\gamma m \dot{x}) = \gamma m \frac {d^2 x}{dt^2} = \gamma_0\beta_0^2 mc^2 \frac{d^2 x}{d(\beta_0 ct)^2} =
   \frac {\gamma_0\beta_0^2 mc^2}{\rho}\,\rho\,\frac{d^2 x}{ds^2} = -qE_{\m x0}\rho\,\frac{d^2 x}{ds^2}  \label {adap2}
\ee    
which combined with Eq. (\ref{f3}) gives rise to
\be 
  \frac{d^2 x}{ds^2} + (3-\beta_0^2-n)\frac{x}{\rho^2} = \frac{d^2 x}{ds^2} + \nu_{\m H}^2\frac{x}{\rho^2} = 0                     \label {eqhor}
\ee
which is just the ``new'' equation for the horizontal betatron oscillations. Similarly we have for the vertical betatron oscillations:

\be
   \frac{d}{dt}(\gamma m \dot{y}) =  \gamma m \frac{d^2 y}{dt^2} = qE_y                                    \label {verteq1}
\ee
where $E_y$ can be found by means of the second Maxwell's equation, in absence of electric charges and expressed in 
cylindrical coordinates; that is
{\small
$$
     \nabla \cdot \vec E = 0 = \frac{1}{R}\frac {\partial}{\partial R}(RE_x) + \frac{\partial E_y}{\partial y} =
              \frac{E_x}{R} + \frac{\partial E_x}{\partial x} + \frac{\partial E_y}{\partial y}
$$ }
or              
{\small          
\ba*
  \frac{\partial E_y}{\partial y} & \simeq & -\frac{E_{\m x0}}{\rho}\left(1-n\frac{n}{\rho}\right)
                     \left(1-\frac{x}{\rho}\right) + n\frac{E_{\m x0}}{\rho} \simeq
            -\frac{E_{\m x0}}{\rho}\left(1-n\frac{n}{\rho}-\frac{x}{\rho}\right) + n\frac{E_{\m x0}}{\rho}   \\  
  \frac{\partial E_y}{\partial y} & \simeq &  (n-1)\frac{E_{\m x0}}{\rho}+(n+1)\frac{E_{\m x0}}{\rho^2} \simeq
                                              (n-1)\frac{E_{\m x0}}{\rho}
\ea* }
Integrating the last expression we obtain $E_y=(n-1)\frac{E_{\m x0}}{\rho}y$, which inserted into Eq. (\ref{verteq1})
generates the equation for the vertical betatron oscillations 

\be
  \frac{d^2 y}{ds^2} + (n-1)\frac{y}{\rho^2} = \frac{d^2 y}{ds^2} + \nu_{\m V}^2\frac{y}{\rho^2} = 0                               \label {eqver}
\ee
having used the same approximations made before. Now, we shall evaluate the value of the field index (\ref{flindex}) in this 
example of cylindrical symmetry, where the radial component of $\vec E$ is
$$
  E_x = \frac{\lambda_q}{2\pi\varepsilon_0}\frac{1}{R} ~~~~~~~~~~~~~ \m{with} ~~~~~~~~~~~~~ \lambda_q = \frac{dq}{ds}
$$
which inserted into the field index (\ref{flindex}) gives rise to
\be
    n = -\rho\frac{2\pi\varepsilon_0}{\lambda}\,\rho
            \left(-\frac{\lambda}{2\pi\varepsilon_0}\frac{1}{\rho^2}\right) = 1              \label {cylflinx}
\ee
which transforms the tunes (\ref{eqhor}) and (\ref{eqver}), relative to a bending sector, into 

\be
        \nu_{\m H}^2 = 2-\beta_0^2 = 1+\frac{1}{\gamma_0^2}  ~~~~~~~~~~ \m{and}  ~~~~~~~~~~
        \nu_{\m V}^2 = 0
\ee
Then the matrices characterizing the horizontal and verical betatron oscillations are
\beqa
   M_{\m H} & = & \left(
\begin{array}{cc}
\cos(\sqrt{2-\beta_0^2}\,\theta) & \displaystyle \frac{\rho}{\sqrt{2-\beta_0^2}}\sin(\sqrt{2-\beta_0^2}\,\theta) \\
-\displaystyle \frac{\sqrt{2-\beta_0^2}}{\rho}\sin(\sqrt{2-\beta_0^2}\,\theta) & \cos(\sqrt{2-\beta_0^2}\,\theta)   \label {MH} 
\end{array} 
\right) \\
M_{\m V}  &=&  \left(
\begin{array}{cc}
1 & \rho\theta \\ 
0 & 1 
\end{array}
\right)              
\eqa
where $\theta=s/\rho$ is the angle subtended by the arc of length $s$.

\section {Lattice and dispersion}
Planning to adopt the usual FODO cell as periodic element, we need to discuss how the beam momentum-spread 
influences the electrostatic optics much more than the traditional magnetic optics. In fact, beyond the natural chromaticity due 
to the lenses, there is the energy variation due to the bending sectors. This effect can be made explicit by making use of 
Eq. (\ref{engain}), namely: 
\be
    -\frac{\beta_0^2}{\rho}x = \frac{\Delta\gamma}{\gamma_0} = \frac{\Delta U}{U_0} = 
    \beta_0^2\frac{\Delta p}{p} ~~~ \Longrightarrow ~~~ \frac{\Delta p}{p} = -\left(\frac{1}{\rho}\right)x =
                  M_{31}x                                  \label {chrombend}    
\ee              
Hence, this new element must be inserted into all the matrices dealing with bending elements. In order to better understanding 
the meaning of modification, let us consider the matrix (2.46) of Ref. \cite{CM}, for simplicity sake.
\be
    M_c = \left(
\begin{array}{ccc} 
\cos(\sqrt{2-\beta_0^2}\,\theta) & \displaystyle \frac{\rho}{\sqrt{2-\beta_0^2}}\sin(\sqrt{2-\beta_0^2}\,\theta) &
          \displaystyle \frac{\rho}{2-\beta_0^2}(1-\cos(\sqrt{2-\beta_0^2}\,\theta))      \\    
- \displaystyle \frac{\sqrt{2-\beta_0^2}}{\rho}\sin(\sqrt{2-\beta_0^2}\,\theta) &  \cos(\sqrt{2-\beta_0^2}\,\theta) &
          \displaystyle \frac{1}{\sqrt{2-\beta_0^2}}\sin(\sqrt{2-\beta_0^2}\,\theta)        \\
- \displaystyle\frac1{\rho} & 0 & 1
\end{array}
\right)                                           \label {newmat}
\ee
Notice that the determinant of the matrix (\ref{newmat}) 
\be
     ||M_c|| = 1 - \left\{\frac {1-\cos(\sqrt{2-\beta_0^2}\,\theta)}{2-\beta_0^2}\right\}        \label {det}
\ee
is not unitarian. This issue is due to the changes of energy  when the particle trajectory is bent. However, after a whole horizontal 
oscillation, particles are accelerated and slowed down in an equal manner meaning that their energy variations are balanced. 
Hence no change of energy takes place and the determinant (\ref{det}) becomes unitarian. In order to better illustrate this topic, 
we consider a perfectly circular machine whose horizontal betatron oscillations wave-length is
$$ 
 \lambda_{\m H} = \frac{2\pi\beta_0 c}{\omega_{\m H}} = \frac{2\pi\beta_0 c}{Q_{\m H}\omega_{\m rev}} = 
 \frac{2\pi\beta_0 c}{\sqrt{2-\beta_0^2}\,\omega_{\m rev}} = \frac{2\pi\rho}{\sqrt{2-\beta_0^2}}
$$
and the angle subtended by a this wave-length is 
$$
   \theta_\lambda = \frac{\lambda_{\m H}}{\rho}= \frac{2\pi}{\sqrt{2-\beta_0^2}} ~~~~~ \m{which ~ yields} ~~~~~
   \cos(\sqrt{2-\beta_0^2}\,\theta_\lambda) = \cos{2\pi} = 1
$$
meaning that the quantity within braces in Eq. (\ref{det}) is null and than the determinant is unitarian. 
As far as the FODO cell is concerned, we insert the element (\ref{chrombend}) into the strip of matrix multiplication shown in 
Eq. {6.17} of Ref.\cite{CM} and we obtain: 
{\footnotesize
$$
 M_{\m H} =  \left(
 \begin{array}{ccc}
1 & 0 & 0 \\
- \displaystyle \frac1{2 f} & 1 & 0 \\
 0 & 0 & 1 
 \end{array}
 \right)\,
 \left(
 \begin{array}{ccc}
1 & \displaystyle \frac{l}{2} & \displaystyle \frac{l\theta_c}8 \\ 
0 & 1 & \displaystyle \frac{\theta_c}2 \\ 
\displaystyle  -\frac1{\rho} & 0 & 1
\end{array}
\right)\,
\left(
\begin{array}{ccc}
1 & 0 & 0 \\ 
\displaystyle \frac1{f} & 1 & 0  \\
0 & 0 & 1
\end{array} 
\right)\,
\left(
\begin{array}{ccc}
1 & \displaystyle \frac{l}2 & \displaystyle \frac{l\theta_c}8 \\
0 & 1 & \displaystyle \frac{\theta_c}2 \\
 \displaystyle -\frac1{\rho} & 0 & 1 
 \end{array}
 \right)\,
 \left(
\begin{array}{ccc}
1 & 0 & 0  \\
\displaystyle -\frac1{2 f} & 1 & 0 \\ 
0 & 0 & 1
\end{array}
\right)
$$
or 
\be
  M_{\m H} = \left(
\begin{array}{ccc}
{\mathcal M}_{11} - \displaystyle \frac{l\theta_c}{8 \rho} & {\mathcal M}_{12} & {\mathcal M}_{13} \\
    {\mathcal M}_{21} - \displaystyle \frac1{2 \rho}\,\left(1- \displaystyle\frac{l}{8 f}\right)\,\theta_c & {\mathcal M}_{22} & {\mathcal M}_{23} \\
\displaystyle -\frac1{\rho}\left(2 - \displaystyle \frac{l}{4 \rho}\right)\,\theta_c & \displaystyle -\frac{l}{2 \rho} & 1 + 
\displaystyle \frac{l\theta_c}{8\rho} 
\end{array}
\right)
\ee
where $\theta_c=\frac{l}{\rho}$ is the azimuth subtended by the FODO cell and the elements ${\mathcal M}$ are just the ones 
of the matrix (6.17) of Ref. \cite{CM}. Hence we obtain
\be 
    \cos\mu_\delta = \frac{1}{2}{\m Tr}M_{\m H} =  
    \frac{1}{2}({\mathcal M}_{11}+{\mathcal M}_{22}) - \frac{l\theta_c}{16\rho} = 
    \cos\mu - \left(\frac{l}{4\rho}\right)^2 
                                                                                     \label {halftrace}
\ee                     
i.e. a slight correction of the phase advance.


\begin {thebibliography} {}

\bibitem {BNL}
AGS Storage Ring EDM Collaboration (Dec. 2009), {\tt www.bnl.gov/edm/review/}

\bibitem {Reiser1}
M. Reiser, {\it Particle Accelerators}, {\bf 4} (1973) 239-247.

\bibitem {Lawson}
J.D. Lawson, {\it The Physics of Charged Particle Beams}, Clarendon Press, Oxford, 1977, pp. 61-69.

\bibitem {Ferrando}
O. Ferrando, Thesis of Degree, Genoa University, 1992.

\bibitem {Reiser2}
M. Reiser, {\it Theory and Design of Charged Particle Beams}, John Wiley $\&$ Sons, Inc., New York,  1994, pp. 111-116.

\bibitem {Luccio}
A.U. Luccio, BNL 16.02.2011 (Draft).

\bibitem {CM}
M. Conte and W.W. MacKay, {\it An Introduction to the Physics of Particle Accelerators}, 2nd ed., World Scientific Singapore (2008).




\end {thebibliography}
\end {document}